\documentclass[journal]{IEEEtran}
\makeatletter
\ifCLASSOPTIONconference
  \def\subsection{\@startsection{subsection}{2}{\z@}%
    {1.5ex plus 1.5ex minus 0.5ex}{0.7ex plus 1ex minus 0ex}%
    {\normalfont\small\centering\scshape}}
\else
  \def\subsection{\@startsection{subsection}{2}{\z@}%
    {3.0ex plus 1.5ex minus 1.5ex}{0.7ex plus 1ex minus 0ex}%
    {\normalfont\small\centering\scshape}}
\fi
\makeatother
\usepackage{diagbox}
\usepackage{graphicx}
\usepackage{subcaption} 
\usepackage{cite}

\usepackage{url}
\usepackage{amsmath,amssymb,amsfonts,mathtools}

\usepackage{array}
\usepackage{amssymb}
\usepackage{xcolor}
\usepackage{verbatim}

\usepackage{multirow}
\usepackage[normalem]{ulem}

\usepackage{soul}
\begin{document}
%
\title{Talk Like a Packet: Rethinking Network Traffic Analysis with Transformer Foundation Models}
%
%
%

\author{{Samara~Mayhoub}, {Chuan~Heng~Foh,~\IEEEmembership{Senior~Member,~IEEE,}} ~{Mahdi~Boloursaz~Mashhadi,~\IEEEmembership{Senior~Member,~IEEE}}, {Mohammad~Shojafar, ~\IEEEmembership{Senior~Member,~IEEE}}, and {Rahim~Tafazolli, ~\IEEEmembership{Fellow,~IEEE}} 
\thanks{S.~Mayhoub is with Aston University, Birmingham, UK (e-mail: mayhoubs@aston.ac.uk).}
\thanks{C.~H.~Foh, M.~B.~Mashhadi, M.~Shojafar and R.~Tafazolli are with 6GIC, Institute for Communication Systems, University of Surrey, Guildford, GU2 7XH, UK (e-mail: \{c.foh, m.boloursazmashhadi, m.shojafar, r.tafazolli\}@surrey.ac.uk).}
 }


\maketitle

\bstctlcite{IEEEexample:BSTcontrol}

\begin{abstract}
Inspired by the success of Transformer-based models in natural language processing, this paper investigates their potential as foundation models for network traffic analysis. We propose a unified pre-training and fine-tuning pipeline for traffic foundation models. Through fine-tuning, we demonstrate the generalizability of the traffic foundation models in various downstream tasks, including traffic classification, traffic characteristic prediction, and traffic generation. We also compare against non-foundation baselines, demonstrating that the foundation-model backbones achieve improved performance. Moreover, we categorize existing models based on their architecture, input modality, and pre-training strategy. Our findings show that these models can effectively learn traffic representations and perform well with limited labeled datasets, highlighting their potential in future intelligent network analysis systems.

\end{abstract}


%
\IEEEpeerreviewmaketitle



\section{{Introduction}}
Transformer-based Large Language Models (LLMs) such as BERT, GPT, LLaMA, and T5 have revolutionized Natural Language Processing (NLP) by achieving high performance in a wide range of tasks, including question answering and text generation. These models learn to understand human language by pre-training on large unlabeled datasets using self-supervised objectives, and then adapting to downstream tasks with minimal fine-tuning.
In networks, an analogous need exists to understand network traffic for \emph{network traffic analysis} tasks such as traffic classification and traffic characteristic prediction.
Historically, these problems were addressed using port-based identification and Deep Packet Inspection (DPI) and later by supervised machine learning (ML) models trained on handcrafted features.
However, increasing encryption, evolving traffic behavior and services, and the high cost of manual labeling have reduced the effectiveness and generalizability of these traditional approaches.

Inspired by the success of Transformer-based LLMs, researchers have begun to explore whether similar models can be applied to a new kind of language: \textit{The Language of Network Traffic}. In this approach, sequences of packets (referred to as flows) are treated like sentences, enabling Transformer-based architectures to model structural and semantic patterns in raw traffic data. This motivates \emph{traffic foundation models}: Transformer backbones pre-trained on large-scale unlabeled traffic and fine-tuned for multiple downstream tasks.

This paper aims to answer the following questions:
Q1) How can Transformer-based foundation models be adapted to learn the \textit{Language of Network Traffic}?
Q2) In what ways do architectural choices, input modalities, and pre-training strategies influence the effectiveness of these models?
Q3) Can such models generalize across diverse downstream tasks to improve performance and support future intelligent network analysis systems?

The main contributions of this article are as follows:

\begin{itemize}
    \item We propose a unified pipeline for pre-training and fine-tuning Transformer-based Traffic Foundation Models.
    \item We present a taxonomy of recent models, categorized by architecture, input modality, and pre-training strategy.
    \item We establish structural awareness as a key design principle for modeling traffic and analyze how models incorporate it across different traffic representation strategies.
    \item We demonstrate that foundation models generalize effectively across three downstream tasks: classification, prediction of traffic characteristics, and generation,  and quantify the benefit of Transformer models via comparisons with non-foundation baselines.
\end{itemize}

The remainder of this paper is organized as follows. We first outline the key challenges in network traffic analysis. We then provide a taxonomy of traffic foundation models. Next, we describe the proposed unified workflows for pre-training and fine-tuning. Through several use-cases, we present fine-tuning results, demonstrate model generalization across diverse downstream tasks and compare with non-foundation baselines. Finally, we outline future directions and conclude the paper.

\section{Challenges in Network Traffic Analysis}\label{sec:challenges}
\textit{Network Traffic Analysis} is essential for network management and cybersecurity, enabling tasks such as traffic classification, intrusion detection, and traffic monitoring~\cite{yatc}. The goal is to extract meaningful insights from raw traffic data, such as identifying applications, detecting attack types, or predicting traffic characteristics like flow volume. However, increasing encryption and emerging services have increased traffic variability, reducing the effectiveness of traditional methods and necessitating more intelligent approaches.
\subsection{Limitations of Traditional Methods}

DPI methods have become ineffective due to widespread encryption, which obscures payloads~\cite{yatc, pert, ET-BERT}. Port-based methods struggle, as applications increasingly use dynamic or shared ports, while DPI fails without access to plaintext payloads. As traffic becomes more complex, these methods lose accuracy, prompting a shift toward learning approaches without payload decryption.

\subsection{Limitations of Supervised Machine Learning Approaches}
Supervised ML has emerged as an alternative to traditional traffic analysis by learning from manually selected flow features~\cite{pert, ET-BERT}. However, early methods often depend on handcrafted statistical features, which lack adaptability to diverse traffic conditions~\cite{lens}. These models also require large volumes of labeled data, which is difficult to obtain, especially due to incomplete handshakes, missing DNS records, or ambiguous server names~\cite{Flow-MAE, pean}. Moreover, supervised ML models often generalize poorly to unseen datasets or zero-day attack types~\cite{trafficllm}, making them challenging to deploy in real-world environments~\cite{qu2024trafficgpt}.

\subsection{Limitations of Supervised Deep Learning Approaches}
Supervised Deep Learning (DL) methods, such as Convolutional Neural Networks, have shown success in modeling complex patterns from raw traffic data~\cite{packetclip}. They eliminate the need for manual feature selection and are widely applied to encrypted traffic classification. However, their performance depends heavily on large, accurately labeled datasets~\cite{lens}, which are often costly or limited by privacy concerns. These models also tend to generalize poorly to unseen traffic or new protocols~\cite{pean}. Additionally, they struggle with capturing long-range dependencies. These challenges have driven interest in self-supervised and pre-training-based methods that leverage raw, unlabeled traffic data.

\subsection{Need for Generalizable, Data-Driven Models}
Modern networks require models that generalize across multiple tasks such as classification, prediction, and generation, while adapting to evolving traffic patterns. As shown earlier, traditional and supervised methods fall short, relying on manual feature engineering and large labeled datasets, which are costly and inflexible. To overcome these limitations, there is a growing shift toward data-driven models that learn directly from raw traffic without handcrafted features. Pre-training and self-supervised learning enable such models to extract rich representations from raw traffic data, reducing labeling costs. These models also support diverse input modalities, making them ideal for unified, multitask traffic analysis~\cite{yatc}. As a result, generalizable, data-driven models are becoming essential components of next-generation network analysis systems.

\section{Transformers as a Foundation for Network Traffic Analysis} \label{sec:transformers}

A Transformer consists of an encoder–decoder architecture built from stacked self-attention and feed-forward layers~\cite{NIPS2017_3f5ee243}. The encoder generates contextualized representations from input sequences, while the decoder produces outputs conditioned on the previously generated tokens and the encoder’s output. Transformer's core innovation is the self-attention mechanism, which lets each token incorporate context from other tokens in the sequence via learned query, key, and value projections; multi-head attention performs this in parallel to capture different types of relationships. In practice, a Transformer can be used as an encoder-only, a decoder-only, or an encoder--decoder backbone, depending on the task.

\section{Foundation Models for Traffic Analysis} \label{sec:taxonomy}

\subsection{Motivation}
The limitations of traditional ML/DL models have prompted a shift toward more data-efficient paradigms. Inspired by the success of Transformers in NLP and Computer Vision (CV), researchers have begun developing Transformer-based foundation models that learn rich traffic representations from unlabeled data. These models serve as backbones for diverse downstream tasks such as classification, flow characteristic prediction, and traffic generation.

\subsection{Key Concepts}
\begin{itemize}
    \item \textit{Self-Supervised Learning (SSL)}: In the context of network traffic, SSL learns meaningful representations from unlabeled traffic by solving pseudo tasks like masked byte prediction, packet ordering, or flow reconstruction. These leverage the inherent traffic structure to produce transferable features.

 \item\textit{Pre-training and Fine-tuning}: SSL is often used for pre-training, where models learn general representations from large unlabeled data. These models are then fine-tuned on smaller labeled datasets for downstream tasks.

\item \textit{Foundation Models for Network Traffic Analysis}: These are large-scale pre-trained models, designed as backbones for diverse traffic analysis tasks. They are often built using SSL on massive traffic datasets, they offer strong generalization on various tasks.
\end{itemize}

\subsection{Taxonomy of Transformer-Based Foundation Models}
We categorize recent foundation models for network traffic analysis into five groups (see Fig.~\ref{fig:taxonomy}), based on their underlying Transformer architecture:
\begin{figure}[!h]
    \centering
    \includegraphics[width=0.4\textwidth]{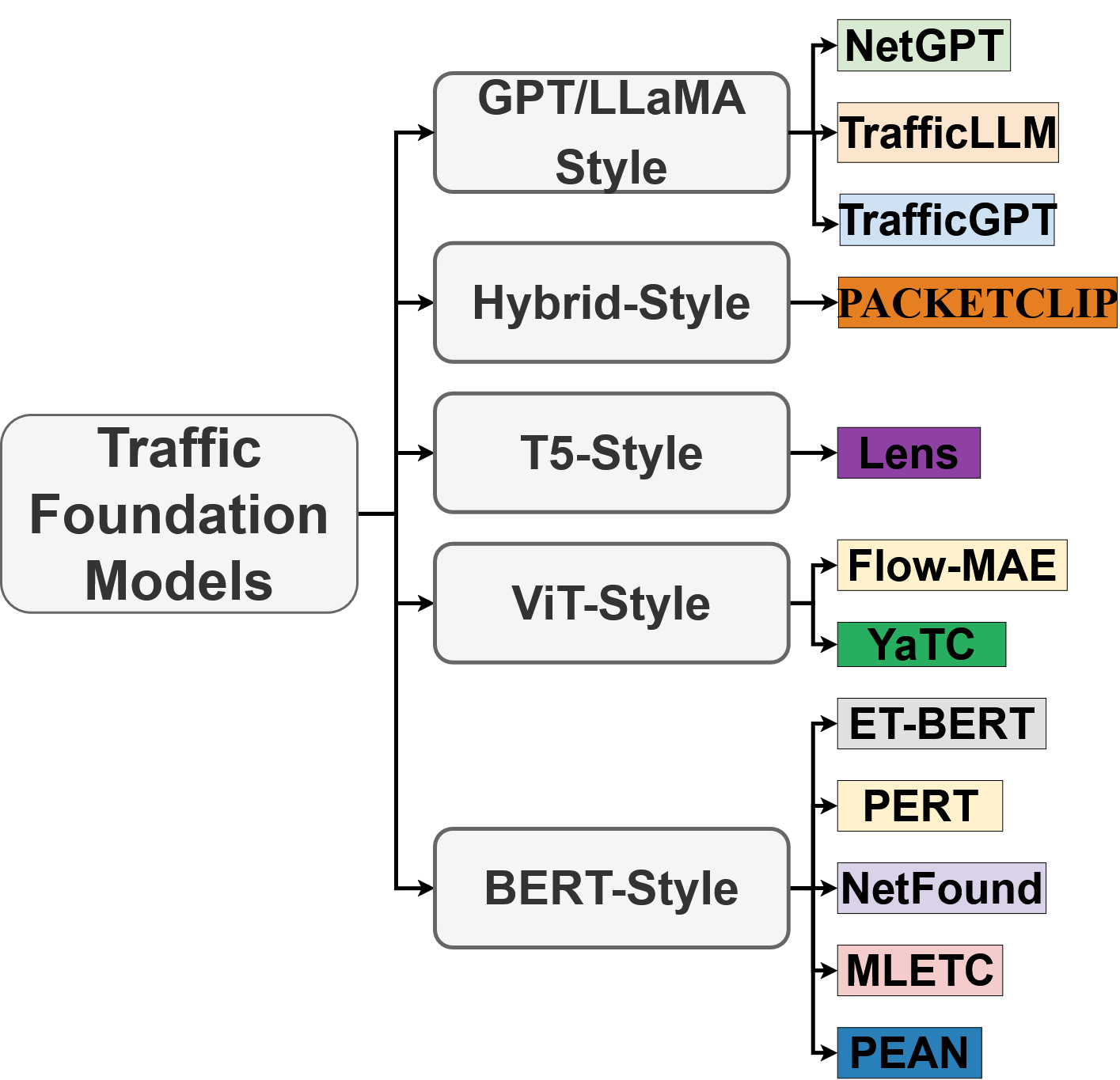}
    \caption{Taxonomy of Transformer-based traffic foundation models.}
    \label{fig:taxonomy}
\end{figure}

\begin{figure*}[!t]
    \centering
    \includegraphics[width=\textwidth]{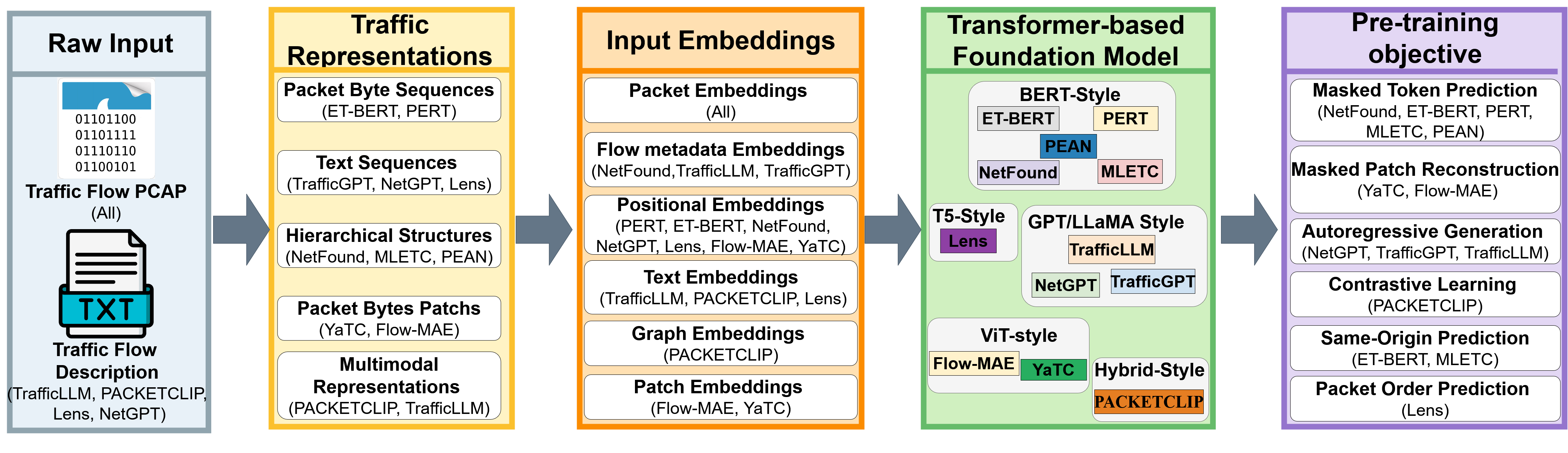}
    \caption{Pre-training workflow for Transformer-based foundation models for network traffic analysis.}
    \label{fig:pretraining_architecture}
\end{figure*}

\subsubsection{Encoder-only Foundation Models (BERT-style)}
These models are inspired by the success of BERT (Bidirectional Encoder Representations from Transformers) in NLP~\cite{bert}. They learn contextualized network traffic representations using encoder-only architectures.

\textit{PERT}~\cite{pert} (Payload Encoding Representation from Transformer) adapts a lightweight BERT to tokenize payload bytes and learn their contextual distributions through pre-training. PERT is motivated by dynamic word embedding in NLP, based on the belief that communication protocols share characteristics with natural human language that can be modeled with NLP-inspired architectures.

\textit{ET-BERT}~\cite{ET-BERT} (Encrypted Traffic Bidirectional Encoder Representations from Transformer) is a BERT-based model designed to learn general representations from unlabeled encrypted traffic. Its key innovation lies in tokenizing raw packets into 'language-like tokens' by extracting BURSTs (unidirectional packet sequences), converting them into hex strings, and applying a bi-gram model with Byte-Pair Encoding. ET-BERT introduces two SSL tasks: Masked BURST Modeling for capturing intra-BURST byte dependencies, and Same-origin BURST Prediction for modeling flow-level context. Compared to PERT, ET-BERT features more advanced tokenization and pre-training strategies.

\textit{PEAN}~\cite{pean} (Packet Embedding Attention Network) uses a BERT-style encoder to capture intra- and inter-packet patterns by combining raw byte sequences with packet length statistics. Pre-training is conducted using a masked language modeling objective.

\textit{netFound}~\cite{netfound} is designed to capture the unique structure and semantics of network data, moving beyond simply treating network traffic as natural language. Its key innovations include: (1) multi-modal embeddings of packet fields and flow metadata; (2) protocol-aware tokenization that segments packets by protocol fields instead of fixed chunks to preserve semantics; and (3) a hierarchical Transformer with skip connections to capture multi-level dependencies. Masked Token Prediction task has been used to pre-train netFound on unlabeled traffic.

\textit{MLETC}~\cite{MLETC} (Multi-Level Encrypted Traffic Classifier) is the first model to explicitly incorporate field-level structure into the pre-training, going beyond byte- and packet-level representations. Built on the DeBERTa architecture, it introduces two SSL strategies: Masked Fields Prediction (MFP) and Same-Origin Flow Prediction (SOFP). MFP masks entire header fields and payload bytes as semantic units, preserving structural integrity. This approach also requires separate models for TCP and UDP due to their distinct header layouts. SOFP, meanwhile, models inter-packet dynamics by predicting whether specific sub-flows originate from the same flow.

\subsubsection{Masked Autoencoder-based Foundation Models (ViT-style)}
These models are inspired by ViT (Vision Transformers), originally developed for CV tasks. They treat network traffic as image-like inputs and often adopt the Masked Autoencoder (MAE) framework for pre-training.

\textit{YaTC}~\cite{yatc} (Yet Another Traffic Classifier) introduces a ViT-based architecture within an MAE framework. It represents traffic using a Multi-level Flow Representation (MFR) matrix, which encodes byte-, packet-, and flow-level information. By treating bytes as pixels, the model captures structured patterns of network traffic. YaTC defines an SSL strategy: Masked Patch Reconstruction, where a high ratio (90\%) of MFR patches are randomly masked, and the model learns to reconstruct the missing content using an encoder–decoder ViT.

\textit{Flow-MAE}~\cite{Flow-MAE} introduces a ViT-based architecture under the MAE framework to improve efficiency in traffic classification. It addresses key limitations of BERT-based models, notably limited input length and reliance on BPE, by replacing it with patch embeddings. Using "bursts" (packets from a flow) as the fundamental unit, Flow-MAE applies a 1D convolutional layer to convert these into fixed-sized patches. It extracts both packet headers and payloads for richer flow representation. Flow-MAE adopts a single SSL task: Masked Patch Modeling, where input patches are randomly masked and reconstructed to learn contextual representations.

\subsubsection{Encoder–Decoder Foundation Models (T5-style)}
These models follow the T5 architecture~\cite{T5}, leveraging an encoder–decoder for both traffic generation and classification tasks.
\textit{Lens}~\cite{lens} applies the T5 framework to network traffic, learning generalized representations via WordPiece-tokenized hex inputs. Lens introduces three SSL tasks: Masked Span Prediction to recover masked spans (contiguous sequence of tokens) and learn contextual representations; Packet Order Prediction to capture temporal information by predicting the order of the first three packets in a flow; and Homologous Traffic Prediction to determine whether two sub-flows originate from the same flow.

\subsubsection{Decoder-only Foundation Models (GPT/LLaMA style)}
These models use autoregressive, decoder-only Transformer architectures (e.g., GPT, LLaMA) to support both traffic generation and understanding via prompting.

\textit{NetGPT}~\cite{netgpt} is a GPT-2-based autoregressive model designed for both traffic generation and classification. It encodes network flows as unified hexadecimal sequences, preserving the semantics of both plaintext and ciphertext traffic. NetGPT reformulates traffic understanding as a generative task and adopts three key strategies: (1) header field shuffling for semantic-preserving augmentation, (2) packet segmentation using special tokens and segment embeddings to retain structure, and (3) prompt-based task conditioning to unify multiple tasks. Pre-training involves next-token prediction over the first three packets of each flow.

\textit{TrafficGPT}~\cite{qu2024trafficgpt} adopts a GPT-style autoregressive architecture enhanced with linear attention to support long sequence modeling. It tokenizes network traffic using hexadecimal representations and encodes temporal information via time interval tokens. To enable the generation of realistic traffic, it introduces a reversible tokenization scheme that allows reconstructing flows back into PCAP format. Pre-training is conducted using a next-token prediction objective over tokenized flows. 

\textit{TrafficLLM}~\cite{trafficllm} is a fine-tuning framework for decoder-only LLMs (such as LLaMA) that enables multimodal learning from text and network traffic via traffic-domain tokenization and instruction-based tuning. It employs a two-stage approach: (1) Natural Language Instruction Tuning to follow expert prompts, and (2) Traffic Tuning for diverse traffic analysis tasks. 

\subsubsection{Hybrid Models}
These models combine Transformer architectures with other neural network components to process multimodal traffic data.
\textit{PACKETCLIP}~~\cite{packetclip} is inspired by Contrastive Language–Image Pre-training (CLIP) technique. It aligns packet payloads with natural language in a shared embedding space using contrastive pre-training and simple projection heads with fixed text and packet encoders (LLaMA and ET-BERT). To enhance semantic reasoning, PACKETCLIP incorporates Graph Neural Networks operating on mission-specific Knowledge graphs generated by LLMs.

\section{Pipeline Overview for Traffic Analysis with Foundation Models} \label{sec:pipeline}

\subsection{Traffic Representation Modalities} The reviewed models adopt diverse strategies to represent raw traffic in formats suitable for Transformers. These representation approaches fall into two categories: 
\subsubsection{Single-Modality}
Models with single-modality rely solely on network traffic data without incorporating external semantic sources like natural language descriptions of the traffic data. 
\textit{Byte Sequences:} It treats network traffic as a continuous stream of raw bytes. For example, PERT~\cite{pert} treats payload bytes as words in a sequence and uses dynamic word embeddings inspired by BERT. Also, ET-BERT~\cite{ET-BERT} converts packet bytes into language-like tokens.
\textit{Hierarchical Structures:} It recognizes that network traffic is not merely a stream of bytes but possesses a multi-layered structure, with information organized at byte, field, packet, and flow levels. For example, MLETC~\cite{MLETC} and NetFound~\cite{netfound} leverage this structure through protocol-aware tokenization, multi-level embeddings, and hierarchical Transformer architectures, enabling understanding across multiple semantic layers of traffic. Similarly, PEAN~\cite{pean} combines raw byte sequences with packet length statistics, modeling intra-packet and inter-packet relationships.
\textit{Image Representation:} It transforms network data into 2D matrices, suitable for computer vision models. For example, YaTC~\cite{yatc} constructs a grayscale matrix per flow. Flow-MAE~\cite{Flow-MAE} also adopts a similar approach by segmenting bursts—treated as 1D byte sequences—into non-overlapping patches using a 1D convolutional layer, which are then embedded for ViT processing.
\textit{Textual Representation:} It converts packet content into hexadecimal strings for processing by LLM tokenizers. For example, Lens~\cite{lens} serializes flow fields into hex-based token sequences with special delimiters to retain structure. Also, NetGPT~\cite{netgpt} and TrafficGPT ~\cite{qu2024trafficgpt} encode each byte of the packet into a hexadecimal number.
\subsubsection{Multi-Modality} These approaches combine different data types such as raw bytes, and natural language, as input.
\textit{Packet + Text:} PACKETCLIP~\cite{packetclip} aligns packet-level data with natural language using contrastive pre-training. It uses ET-BERT to tokenize packets and LLaMA to generate descriptions from flow data. On the other hand, TrafficLLM~\cite{trafficllm} combines input derived from raw traffic (extracted using TShark) with expert-crafted natural language instructions. 

\begin{figure*}[!t]
  \centering
  \includegraphics[width=0.95\linewidth]{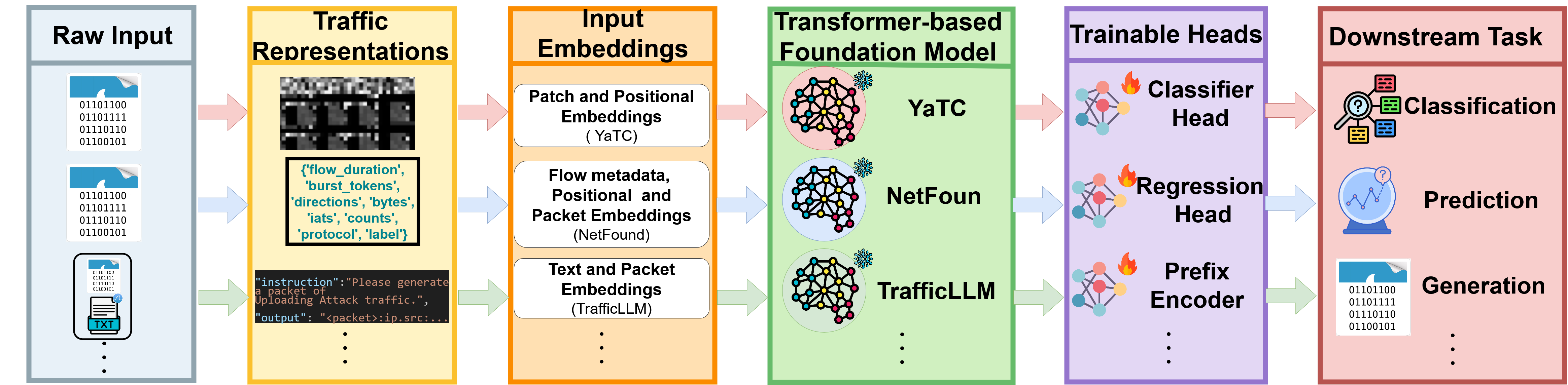}
  \caption{Fine-tuning workflow for various network traffic analysis tasks.}
  \label{fig:foundation-models-tasks}
\end{figure*}

\subsection{Encoding Structure Awareness in Network Traffic}

Transformers are inherently permutation-invariant, meaning they do not understand the order of input tokens. Therefore, auxiliary mechanisms are necessary to capture sequence order and structural information. In network traffic, structural awareness is crucial due to the hierarchical nature of flows: a flow consists of multiple packets, each comprising a header and a payload, with headers containing structured protocol fields. Structural awareness is addressed through a range of modeling approaches:
\subsubsection{\textit{Standard Positional Embedding}} It adds fixed or learned vectors to input tokens to provide the model with information about token order. This approach is a common NLP technique adapted in traffic models like ET-BERT~\cite{ET-BERT}, PERT~\cite{pert}, NetGPT~\cite{netgpt}, and NetFound~\cite{netfound}.
\subsubsection{\textit{Special Time and Semantic Tokens}} uses special tokens like \texttt{[start]} and \texttt{[time]} to represent packet boundaries and time intervals, encoding temporal and structural semantics in inputs. Lens~\cite{lens} also uses special tokens like \texttt{[pkt]} to mark packet boundaries and \texttt{[head]} to separate header fields from payloads. PEAN~\cite{pean} uses a \texttt{[PACKET]} token to aggregate byte-level information into a packet-level representation. ET-BERT~\cite{ET-BERT} differentiates sub-bursts using a [SEP] special token and segment embeddings. TrafficLLM~\cite{trafficllm} uses \texttt{[packet]} as an indicator token to signify the beginning of traffic data. MLETC~\cite{MLETC} tokenizes traffic at byte, field, and packet levels using structural markers like \texttt{[PKT]} and \texttt{[HDR]} to denote boundaries, and employs header embeddings to distinguish protocol fields, enabling semantic understanding across protocol layers. NetGPT~\cite{netgpt} appends \texttt{[pck]} to each packet as a segment delimiter. TrafficGPT~\cite{qu2024trafficgpt} introduces \textit{Time Interval Tokens} that encode inter-packet delays in exponential format, enriching temporal structure awareness.
\subsubsection{\textit{Hierarchical Embeddings}} Beyond simple token order, several models explicitly capture the inherent hierarchical structure of network traffic. For example, YaTC~\cite{yatc} captures multi-level structure by encoding traffic into a matrix with dedicated regions for byte-, packet-, and flow-level features. Moreover, Flow-MAE~\cite{Flow-MAE} captures structural awareness by processing traffic bursts as one-dimensional sequences that preserve packet structure.
\subsubsection{\textit{Graph Embeddings}} PACKETCLIP~\cite{packetclip} incorporates knowledge graph structures generated from natural language inputs to capture domain-specific semantics across flows and packets, enabling GNN-based processing alongside Transformer embeddings.

\subsection{Pre-training Strategies}

Pre-training is crucial for building generalizable and scalable traffic models. Inspired by SSL in NLP and CV, foundation models adopt diverse pre-training strategies, Fig.~\ref{fig:pretraining_architecture}

    \subsubsection{\textit{Masked Language Modeling}} Inspired by BERT, this strategy hides parts of the input and trains the model to predict them, enabling learning of contextual and semantic representations. Variants of this approach are used in PERT~\cite{pert}, ET-BERT~\cite{ET-BERT}, MLETC~\cite{MLETC}, PEAN~\cite{pean} and NetFound~\cite{netfound}. 
    
    \subsubsection{\textit{Masked Image Modeling}} Inspired by MAE, this strategy masks portions of image-like traffic inputs and reconstructs them to learn spatial and structural patterns. YaTC~\cite{yatc} and Flow-MAE~\cite{Flow-MAE} follow this approach.

    \subsubsection{\textit{Contrastive Learning}} This approach trains the model to align matching representations from different modalities in a shared embedding space. PACKETCLIP~\cite{packetclip} applies this to link packet-level features with their descriptions.
    
    \subsubsection{\textit{Same-Origin Prediction}} This strategy trains the model to determine whether different traffic segments belong to the same flow, capturing flow-level dependencies. ET-BERT~\cite{ET-BERT} and MLETC~\cite{MLETC} apply variants of this strategy.
    
    \subsubsection{\textit{Packet Order Prediction}} Introduced by Lens~\cite{lens}, this strategy trains the model to reorder shuffled packets within a flow, reinforcing temporal coherence.
    
    \subsubsection{\textit{Generative Pre-training}} This strategy models traffic as a sequence of tokens and trains the model to generate the next token, learning sequential and generative patterns. NetGPT~\cite{netgpt}, TrafficLLM~\cite{trafficllm} and TrafficGPT~\cite{qu2024trafficgpt} apply this approach.

These pre-training strategies equip models with a deep understanding of traffic structure and semantics. Fine-tuning then adapts these models to downstream tasks

\subsection{Fine-tuning on Downstream Tasks}

Various foundation models can support downstream tasks, Fig.~\ref{fig:foundation-models-tasks}. These tasks can be categorized into the following three general groups:

\subsubsection{ \textit{Traffic Classification}} This remains the most widely supported downstream task. This includes tasks such as Encrypted Application and Service Classification, Malicious Traffic and Intrusion Detection, Attack Type Classification and VPN Detection. Models like NetFound~\cite{netfound}, Flow-MAE~\cite{Flow-MAE}, ET-BERT~\cite{ET-BERT}, MLETC~\cite{MLETC}, PERT~\cite{pert}, and PACKETCLIP~\cite{packetclip} have been specifically fine-tuned for traffic classification tasks.
\subsubsection{ \textit{Traffic Generation}} This emerging task focuses on synthesizing realistic network traffic at different levels, including header field generation (e.g., IP addresses, ports), full packet synthesis, and protocol-specific traffic. Models such as Lens~\cite{lens}, NetGPT~\cite{netgpt}, TrafficGPT~\cite{qu2024trafficgpt}, and TrafficLLM~\cite{trafficllm} support these capabilities. Traffic Generation task is useful in security testing, network simulation, traffic data augmentation (especially when data is sparse or imbalanced), and \textit{network digital twins}.

\subsubsection{\textit{Traffic Characteristic Prediction}} Although none of the reviewed models were explicitly fine-tuned for prediction tasks, models such as NetFound~\cite{netfound} exhibit the potential to be adapted for predicting traffic characteristics. These include parameters like packet length, inter-arrival time, traffic volume, and throughput. Traffic characteristic prediction tasks are useful in network management and cybersecurity applications. 
\subsection{Datasets}
\label{subsec:datasets_eval}
Datasets for traffic analysis differ in realism, application/protocol coverage, and release format. For \emph{pre-training} traffic foundation models, datasets with real, raw PCAP traces are particularly valuable because they preserve packet- and flow-level structure and reflect real traffic dynamics. For \emph{fine-tuning}, labeled PCAP-based datasets are preferred because they provide task-specific ground truth for supervised adaptation. Dataset selection is also commonly guided by FAIR principles (Findable, Accessible, Interoperable, Reusable), favoring datasets that are publicly available and well-documented, while also accounting for \emph{privacy} constraints through appropriate prepossessing (e.g., anonymizing IP addresses).

\section{Use Cases: Generalizability and Fine-tuning for Downstream Tasks} \label{sec:downstream}
To assess generalization, we fine-tuned three Transformer-based network foundation models on two previously unseen datasets and evaluated them on three key downstream tasks, demonstrating their generalizability to unfamiliar network environments.

We use two public datasets: \textit{CICIoT2023}\footnote{\url{https://www.unb.ca/cic/datasets/iotdataset-2023.html}}, which provides raw PCAP traces of IoT benign traffic and a broad set of IoT-focused attacks, and \textit{CIC-IDS-2017}\footnote{\url{https://www.unb.ca/cic/datasets/ids-2017.html}}, which captures five days of benign and attack traffic from a simulated testbed and is released as raw PCAPs and bidirectional flow records.

\subsection{Generalization Across Traffic Classification Tasks}
To demonstrate the generalization capabilities of Transformer-based foundation models in traffic classification scenarios for IoT network security applications, we fine-tuned YaTC, as shown in Fig.~\ref {fig:foundation-models-tasks}, on CICIoT2023 dataset. While the full dataset includes a broader range of IoT-related attacks, we selected 16 attacks and benign traffic. As CICIoT2023 dataset provides raw PCAP files, we first segmented the traffic into flows and then extracted the packet bytes to construct grayscale images. These images were used for fine-tuning YaTC with a classifier head (a multi-layer perceptron (MLP) with three hidden layers and a softmax output layer).
The fine-tuned YaTC encoder with classifier head achieved an overall classification accuracy of 96.9\%. The classifier head by itself achieved an overall accuracy of 72.5\%. The model demonstrated consistent performance across all attack types, with per-class F1-scores exceeding 0.93 for most categories. Detailed performance metrics per class are summarized in Table~\ref{tab:yatc-vs-mlp} for both YaTC with classifier head and the MLP-only baseline, and clearly shows the superiority of using a Transformer-based foundation model to capture richer traffic structure and semantics for IoT attack classification.

\begin{table}[!t]
\centering
\caption{Classification performance of fine-tuned YaTC with classifier head vs. classifier head only.}
\label{tab:yatc-vs-mlp}

\scriptsize
\setlength{\tabcolsep}{1.8pt}
\renewcommand{\arraystretch}{1.05}

\resizebox{0.95\columnwidth}{!}{%
\begin{tabular}{|l|ccc|ccc|}
\hline
\multirow{2}{*}{\textbf{Attack Class}} &
\multicolumn{3}{c|}{\textbf{YaTC+MLP}} &
\multicolumn{3}{c|}{\textbf{MLP}} \\
\cline{2-7}
& \textbf{Precision} & \textbf{Recall} & \textbf{F1}
& \textbf{Precision} & \textbf{Recall} & \textbf{F1} \\
\hline
Benign                 & 0.9950 & 0.9910 & 0.9930 & 0.4802 & 0.7870 & 0.5965 \\
DDoS HTTP Flood        & 0.9885 & 0.9845 & 0.9865 & 0.9145 & 0.7965 & 0.8514 \\
ACK Fragmentation      & 0.9900 & 0.9905 & 0.9903 & 0.9950 & 0.9920 & 0.9935 \\
Vulnerability Scan     & 0.9786 & 0.9850 & 0.9818 & 0.3423 & 0.3900 & 0.3646 \\
SQL Injection          & 0.9612 & 0.9041 & 0.9318 & 0.5440 & 0.1001 & 0.1691 \\
Port Scan              & 0.9806 & 0.9840 & 0.9823 & 0.5147 & 0.3675 & 0.4288 \\
OS Scan                & 0.9819 & 0.9745 & 0.9782 & 0.4196 & 0.0470 & 0.0845 \\
Host Discovery         & 0.9940 & 0.9925 & 0.9932 & 0.6545 & 0.6840 & 0.6689 \\
Mirai UDP Flood        & 0.8632 & 0.8873 & 0.8751 & 0.9990 & 0.9930 & 0.9960 \\
Mirai-greeth Flood     & 0.8689 & 0.8167 & 0.8420 & 0.9949 & 0.9820 & 0.9884 \\
Mirai-greip Flood      & 0.9539 & 0.9673 & 0.9606 & 0.9856 & 0.9955 & 0.9905 \\
MITM-ARP Spoofing      & 0.9705 & 0.9855 & 0.9779 & 0.6557 & 0.6705 & 0.6630 \\
DNS Spoofing           & 0.9790 & 0.9785 & 0.9787 & 0.6840 & 0.6765 & 0.6802 \\
Dictionary-BruteForce  & 0.9145 & 0.9408 & 0.9274 & 0.5245 & 0.5935 & 0.5569 \\
DDoS-UDP Fragmentation & 0.9472 & 0.9442 & 0.9457 & 0.9930 & 0.9885 & 0.9907 \\
DDoS-Slowloris         & 0.9875 & 0.9905 & 0.9890 & 0.9552 & 0.9810 & 0.9679 \\
DoS HTTP Flood         & 0.9890 & 0.9920 & 0.9905 & 0.8261 & 0.9215 & 0.8712 \\
\hline
\textbf{Average} & 0.9614 & 0.9593 & 0.9602 & 0.7343 & 0.7039 & 0.6978 \\
\hline
\end{tabular}}
\end{table}

\subsection{Generalization across Traffic Characteristic Prediction Tasks}
To evaluate the ability of traffic foundation models to perform traffic volume prediction, we fine-tuned NetFound, as shown in Fig.~\ref{fig:foundation-models-tasks}, on the benign subset of CIC-IDS-2017 dataset. Each flow was extracted from raw PCAP files, tokenized into burst-level payloads, headers, and metadata (packet size, inter-arrival time, direction, etc.), and labeled with its total transmitted volume in bytes, forming the dataset. The dataset was used for fine-tuning NetFound with a regression head (MLP with three hidden layers and LeakyReLU activations).
The fine-tuned model demonstrated strong predictive performance, achieving a low Mean Absolute Percentage Error (0.082\%), high $R^2$ score (0.934), and low Mean Absolute Error (MAE) of 18 bytes. On the other hand, the baseline MLP achieved an MAE of 21.63 bytes and an $R^2$ of 0.840. These results highlight the model’s ability to generalize to network traffic characteristic prediction tasks.

\begin{figure}[!h]
    \centering
    \includegraphics[width=0.5\textwidth]{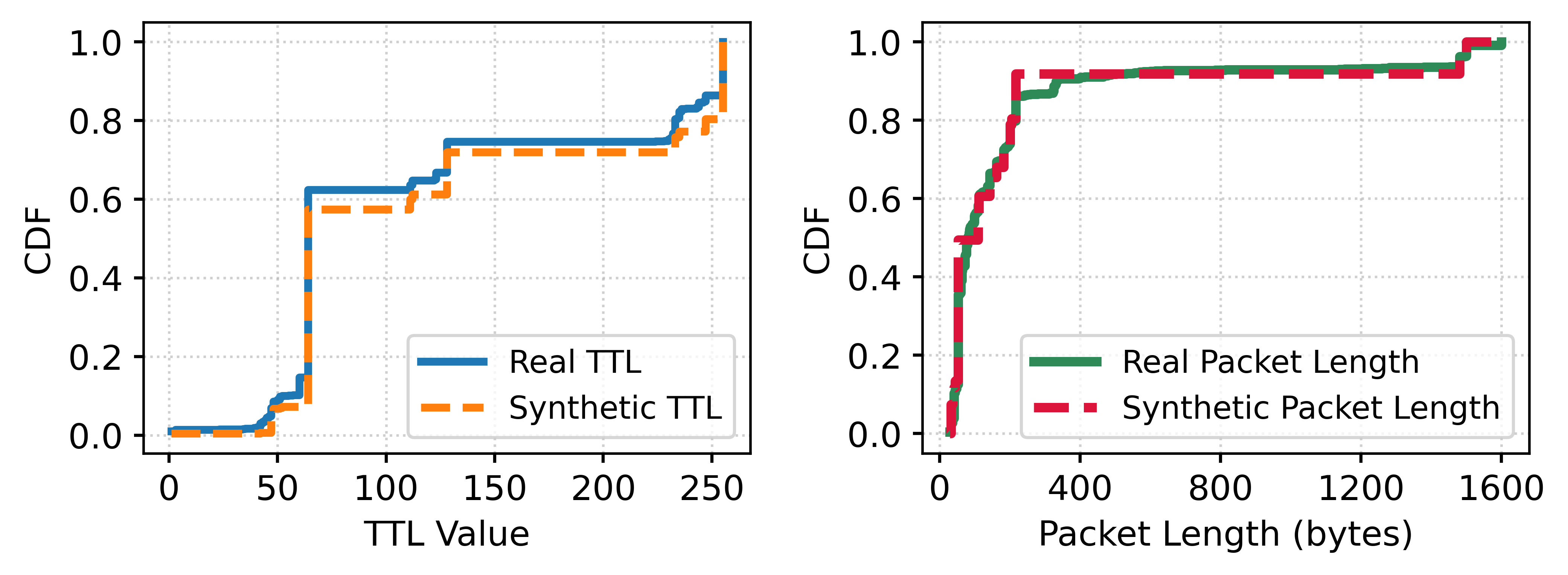}
    \caption{CDF of TTL and packet length for real vs. TrafficLLM-generated packets.}
    \label{fig:ttl-cdf-comparison}
\end{figure}

\subsection{Generalization across Traffic Generation Tasks}
To assess the capability of traffic foundation models in generation tasks, especially when training data is sparse or imbalanced, we fine-tuned TrafficLLM on a PCAP trace of Uploading Attack flows from the CICIoT2023 dataset. This attack involves unauthorized file uploads and is underrepresented in the dataset, making it a suitable candidate for synthetic augmentation. We used \texttt{tshark} to extract packet-level summaries, which were reformatted into instruction–output pairs as shown in Fig.~\ref{fig:foundation-models-tasks}.
Using P-Tuning v2, we fine-tuned a prefix encoder on top of a frozen ChatGLM2-6B model to autoregressively generate packet-level outputs conditioned on prompts. We then compared the distributions of two packet fields—Time-To-Live (TTL) and IP packet length—between real and generated traffic using Cumulative Distribution Functions (CDFs), as shown in Fig.~\ref{fig:ttl-cdf-comparison}. The close alignment in both plots indicates that the fine-tuned model effectively replicates key
packet fields of real network traffic.

\section{Future Research Directions for Traffic Foundation models}\label{sec:future}

While Transformer-based foundation models have shown promise for network traffic analysis, several open research directions remain to realize their full potential.

\subsection{Computational Complexity and Real-Time Constraints}
Transformer architectures have a quadratic time and memory complexity with respect to input sequence length, $O(n^2)$. This becomes challenging when longer packet flows/bursts are processed. Future work can mitigate this overhead by adopting more efficient attention mechanisms (e.g., sparse or low-rank approximations).
In addition, techniques such as knowledge distillation, parameter pruning, and model quantization can be applied to deploy these models on less efficient hardware.

\subsection{Latency-performance trade-offs} In practice, network monitoring systems require sub-second inference. Future work should benchmark not just accuracy but the end-to-end inference latency on the target hardware. Speculative inference, computation offloading, and hierarchical processing pipelines can achieve a better accuracy/latency trade-off.

\subsection{Explainability and Trust}
To gain operational trust, more explainable traffic foundation models are required. Explainable AI and LLM models can be applied to provide human-readable explanations in real-time from the Transformer attention weights and structural embeddings provided by the traffic foundation model. These can attribute the model predictions to specific traffic components (e.g., header fields, temporal bursts). This is especially critical for security-sensitive applications.

\subsection{Reasoning-Enhanced Traffic Analysis}

Future traffic foundation models may support lightweight evidence-based reasoning over packet/flow structure and time, and provide short human-readable justifications, improving robustness under distribution shifts, partial data, and attacks.

\section{Conclusion}\label{sec:conclusion}
In this paper, we introduced the pre-training and fine-tuning pipeline for Transformer-based traffic foundation models, which can generalize to various categories of downstream tasks. Inspired by advances in NLP, these models treat traffic as a learnable language and leverage self-supervised pre-training to build generalizable representations. We explored various Transformer foundation models for traffic classification, flow characteristic prediction and traffic generation tasks. We categorized the existing models based on their Transformer architecture (e.g., encoder-only, decoder-only, encoder–decoder), input modalities (e.g., packets, flows, images, text), and pre-training strategies (e.g., masked modeling, contrastive learning). Our experimental results further demonstrate the ability of Transformer foundation models to generalize across these tasks. We also compare against non-foundation baselines demonstrating that the foundation-model backbones achieve improved performance.
These findings highlight the potential of foundation models for robust, scalable, and data-efficient traffic analysis.

\ifCLASSOPTIONcaptionsoff
  \newpage
\fi


\bibliographystyle{IEEEtran}
\bibliography{IEEE_references}

%


\section*{Biographies}
\vskip -3\baselineskip plus -0.9fil
\begin{IEEEbiographynophoto}{{Samara Mayhoub}}
is a Lecturer in Cyber Security at Aston University. She held two postdoc positions: University of Surrey (6GIC, 2024) and Queen’s University Belfast (CSIT, 2025). She received her Ph.D. in Network Security from Samara National Research University in 2022.
\end{IEEEbiographynophoto}

\vskip -3\baselineskip plus -0.9fil
\begin{IEEEbiographynophoto}
{Chuan Heng Foh} (Senior Member, IEEE) is an Associate Professor in the 6G Innovation Centre at UoS, the Vice Chair of the IEEE VTS Ad Hoc Committee on Mission Critical Communications. He is on the editorial boards of several international journals and a Senior Editor of IEEE ACCESS.

\end{IEEEbiographynophoto}

\vskip -3\baselineskip plus -0.9fil

\begin{IEEEbiographynophoto}{Mahdi Boloursaz Mashhadi}(Senior Member, IEEE) is a Lecturer at the 6G Innovation Centre, UoS, and a Surrey AI fellow. His research is focused at the intersection of AI/ML with wireless communications. He is a PI/Co-PI for various government and industry-funded projects including the UKTIN/DSIT 12M£ national project TUDOR. 
\end{IEEEbiographynophoto}

\vskip -3\baselineskip plus -0.9fil

\begin{IEEEbiographynophoto}{Mohammad Shojafar}
(Senior Member, IEEE) is an Associate Professor in the 6G Innovation Centre at UoS, Marie Curie Alumni, an Intel Innovator and Surrey Sustainability Fellow. He has secured more than £2M in research funding as a PI for various EU- and UK-funded projects. He is also a member of the ETSI Intelligent Transportation Systems Group, GSMA Open-Telco LLM Group, and the 3GPP 5G SSA working group.
\end{IEEEbiographynophoto}

\vskip -3\baselineskip plus -0.9fil
\begin{IEEEbiographynophoto}
{Rahim Tafazolli } (Fellow, IEEE) is a Regius Professor of Electronic Engineering, a Professor of Mobile and Satellite Communications, and the Founder and the Director of 5GIC, 6GIC, and the Institute for Communication Systems, UoS. He has more than 30 years of experience in digital communications research and teaching. 

\end{IEEEbiographynophoto}





\end{document}